\PassOptionsToPackage{cmyk,table}{xcolor}
\documentclass[conference,flushend]{iaria}
\pdfoutput=1 

\usepackage[babel=true,english=american]{csquotes}
\usepackage[english]{babel} 

\usepackage[shortcuts]{extdash} 

\usepackage[alwaysadjust]{paralist}
\usepackage[caption=false,font=footnotesize]{subfig}
\usepackage{comment}
\usepackage{xcolor}
\usepackage{soul}

\usepackage[
babel=true, 
expansion=alltext,
protrusion=alltext-nott, 
nopatch=eqnum, 
final 
]{microtype}

\usepackage{fontawesome} 
\usepackage{relsize}     
\usepackage{lipsum}      

\title{C-LISTEN: Cognitive Load Impacts of Sensory-Triggered Environmental Navigation in Virtual Reality}

\author{
    \IEEEauthorblockN{
        Md. Monowar Hossain\,\orcidlink{0000-0001-9317-8980}
    }
    \IEEEauthorblockA{
        Department of Computer Science\\
        Kennesaw State University, United States\\
        e-mail: \texttt{murad0904045@gmail.com}\\
\texttt{mhossa36@students.kennesaw.edu}    }

    \and

    \IEEEauthorblockN{
        M. Rasel Mahmud
    }
    \IEEEauthorblockA{
        Assistant Professor of Computer Science\\ 
        Kennesaw State University, United States\\
        e-mail: \texttt{m.raselmahmud1@gmail.com}\\
\texttt{mmahmud2@kennesaw.edu}    }
}

\begin{document}

\maketitle

\begin{abstract}
Nowadays, Virtual Reality (VR) systems are more helpful for making decisions, training, and rehabilitation. Multimodal settings in these systems impose a significant balance and cognitive difficulties. Additionally, decreased performance, higher cognitive load, user overwhelm, and limited accessibility of VR technology can be caused by excessive auditory, visual, and sensory stimulation. In our pilot study, we examine the strategic design of directional auditory cues to improve cognitive load and enhance user experience in virtual environments. This research also enhances fundamental knowledge that auditory feedback reduces cognitive load in VR, with statistical analysis confirming this improvement (p = .0011). This study investigates how postural balance and pupil diameter correlate with cognitive load while the users navigate. Moreover, this study provides realistic overview design guidelines for accessibility, which will make VR experiences safer and more useful for everyone, including people with cognitive impairments.
\end{abstract}


\section{Introduction}
\label{sec:introduction}
Virtual Reality (VR) has the ability to create interactive and immersive environments that are used in different sectors, such as healthcare, training, education, and entertainment. Cognitive load refers to the amount of mental effort for processing information while accomplishing a task, which is strongly influenced by the effectiveness of VR experiences. One of the major reasons for ineffective design is that it leads to a degradation of performance, discomfort, and overload of users’ cognitive resources.

The cognitive load is related to the participants’ pupil diameter and body balance. Kahneman et al. \cite{kahneman1966pupil} found that when a participant processed information in short-term memory, the pupil diameter was larger than in a normal situation. Pupil size increased when performing a difficult task, which reflects the amount of mental effort or cognitive load on memory. The impact continues even when visual factors remain controlled, which indicates that pupil dilation is associated with cognitive processing rather than only adjustments in eye focus. In general, the size of the pupil is a good predictor of how much memory and mental work a person is utilizing at any given time.

In Small et al.'s work \cite{small2021influence}, participant body balance is degraded during the performance of a listening or spelling task involving short and long words, as well as the walking dual task condition. The researchers established that the participant prioritized the thinking task over balance, and a higher cognitive load led to a decline in balance control during walking. In \cite{mahmud2024multimodal}, without a feedback system in immersive VR, the standing balance generally declined, and adding assistive feedback improved the balance. They noticed that individual assistive feedback (auditory, vibrotactile, or visual) can improve balance without noticeably increasing cognitive load than the multimodal condition.

We investigated 15 types of audio, categorized into
four types (classical, natural, electronic, and unnatural), which
helped to reduce cognitive load. In the classical music conditions included four common music from Western classical music: Adagio in G minor (Albinoni), Allegro con spirito from Sonata K.448 (Mozart), Gymnopédie No. 1 (Satie), and Für Elise (Beethoven). These musics were chosen because they are common for listening conditions in cognitive research \cite{barza2025mozart, borella2014age, xing2016mozart, jauvsovec2005influence, shen2023effect, mohammadian2025effect}. The natural soundscape had a common combination of sounds, including fountains and birds tweeting, ocean waves, and birdsong. Researchers often use them to examine how they can help people feel less stressed and less mentally burdened \cite{alvarsson2010stress, medvedev2015restorative}. High-energy tracks, such as as Balearic Pumping, Natural (Imagine Dragons), Rum n’ Bass (Boom Kitty), and POP/STARS (K/DA) are known as electronic music. Strong rhythmic drive, amplified bass, and fast, repetitive beats are features included in this type of music that affect arousal and attentional activation. During the task, these were used to investigate how they may influence focus and cognitive load \cite{ellahiyoun2025effects}. There are commonly unnatural sounds like aviation noise, motorcycle noise, white noise, and construction noise used that represent the human-made acoustic environment in research to examine how they can impact stress and cognitive load \cite{medvedev2015restorative, helps2014different, mahmud2024multimodal}.

The key contributions of our research include: 
First, this study investigates whether assistive auditory cues can help users manage cognitive load in virtual reality (VR) environments. The results show that well-designed auditory guidance can help users control their mental effort during VR navigation tasks. Second, the study analyzes how different categories of audio cues affect cognitive load. This highlights the importance of selecting appropriate auditory feedback when designing VR systems. Third, this work examines the relationship between postural balance and cognitive load. The findings suggest that higher cognitive load can affect users’ physical stability during VR interaction. Finally, the study identifies a correlation between pupil diameter and cognitive load, indicating that pupil dilation can be used as a physiological measure to monitor users’ mental effort in immersive environments. Overall, these findings provide useful insights into how auditory feedback influences both cognitive and physical responses in VR, which can help in designing safer and more efficient VR systems.

The rest of the paper is explained as follows: In Section \ref{sec:background_study}, a background study is discussed, which contains the previous research summary of the assistive auditory feedback system done by other researchers. In Section \ref{sec:methods}, we describe our proposed methods. Lastly, Sections \ref{sec:stat_analysis}, \ref{sec:discussion}, and \ref{sec:conclusion} detail the statistical analysis \& experimental results with discussion and conclusion, respectively.

\section{Background Study}
\label{sec:background_study}

This section reviews existing literature on assistive auditory feedback systems and their role in influencing cognitive load, user performance, and navigation in virtual environments.

\subsection{Assistive Auditory Feedback Technology}
While improving the efficiency of task completion, controlling cognitive load is one of the vital goals of assistive auditory feedback technology, which provides information through audio. Well-designed auditory feedback cues, such as earcons, auditory icons, speech prompts, and sonification, can enhance user performance and support usability, which is related to reducing unnecessary cognitive effort \cite{peres2008auditory}. 

Multimodal feedback can reduce task difficulty in VR tasks \cite{skarbez2017survey, hossain2025too}. To find out how VR-based assistive auditory, visual, vibrotactile, or multimodal feed-back can help people with balance and gait problems in \cite{mahmud2023auditory}. Furthermore, Mahmud et al. \cite{mahmud2024multimodal} found that cognitive load is higher (p = .04) when multimodal feedback is used than when individual feedback is used. Ricci et al. also found that a single modality, like auditory or haptic cues, is more effective than multimodal signals due to helping to easily distribute attention in complex situations for navigation.  Auditory feedback can enhance learning and memory utilization by reinforcing task-relevant cues \cite{brickman2000multisensory} and improve postural balance \cite{mahmud2022auditory}. Real-world studies demonstrate that auditory assistive cues can improve navigation, attention management, and error correction \cite{scharine2009auditory, mahmud2022auditory}.

Although complex sound can affect the cognitive load for the navigation system, for those who have a problem with visualization, the auditory feedback helps to navigate through detecting obstacles and plotting a path \cite{ricci2024navigation}. For a safety-critical environment like automated driving, the "speech" and “spearcon” cues performed better than earcons, as they helped to understand the critical situation faster and solve the situation using less mental stress \cite{chai2024effects}. Bruckman et. al. \cite{vargas2021auditory} analysed that helps cognitive support in IDEs using the auditory feedback. This auditory feedback provides information on code syntax and error details in programming tasks for solving problems.

\subsubsection{Classical Music as an Auditory Feedback}
Mozart’s Sonata is the most popular piece used for research in music, cognitive and brain activity. The authors examined the past 30 years of studies on humans and animals that revealed how the person’s excitement, preferences, and listening context affected their memory, attention, and brain activity \cite{barza2025mozart}. Mozart's Sonata and Adagio in G Minor music were used in the study \cite{borella2014age}, and the researcher examined how these influenced the working memory in different age groups. They also found that there is no overall effect on complex tasks. In Xing et al.'s work \cite{xing2016mozart}, for improving learning and task performance, the music of "Mozart’s Sonata" is a better solution than other  versions that have increased performance and provide a more positive effect on brain activity. 

In Jauvsovec et al.'s study \cite{jauvsovec2005influence}, while solving spatial rotation tasks and math tasks, the brain activity is affected by listening to Mozart’s music. This music degraded the performance of the math task, but it provided a better performance in the spatial task. The researchers analyzed that the music can boost the activity in the brain area that assists in combining the information into meaningful. Shen et al. \cite{shen2023effect} investigated how the soft background music (Gymnopedie No.1) without lyrics affects participants' attention and showed that this music performs better than silence. They also suggested that soft background music can improve participant attention and quick learning. Mohammadian et al. \cite{mohammadian2025effect} explored how listening to “Für Elise” music in noisy simulated open-plan offices contributed to improving the mental workload more than using irrelevant speech. They used n-back tests for assessing performance accuracy and provided suggestions that the background music can act as a counteraction to noise and provide more support for cognitive performance. We chose auditory feedback over vibrotactile \cite{mahmud2025vibrotactile, mahmud2022vibrotactile, mahmud2022standing} or visual \cite{mahmud2023visual, mahmud2023eyes} feedback because participants had to wear vibrating devices for vibrotactile feedback and had to look at visual signs for visual feedback, which may increase their cognitive load. 
  
\subsubsection{Natural Music as an Auditory Feedback}
The author \cite{alvarsson2010stress} examined that a combination of sounds from a fountain and tweeting birds reduces the stress level than noisy sounds. Moreover, nature sounds can restore skin conductance levels more quickly after completing the stressful task. They also observed that for participants' relaxation, natural sounds play a vital role. After stressing and during rest, they \cite{medvedev2015restorative} analyzed how different natural sounds, such as "ocean waves and birds’ songs" affect the participants’ mental workload. They also found that heart rate decreases after completing a stress task and that skin conductance increases when participants hear natural sounds during rest, helping reduce cognitive load.

\subsubsection{Electronic Music as an Auditory Feedback}
The authors investigated \cite{ellahiyoun2025effects} how musical training affected users’ cognitive load and task performance during VR Beat Saber gameplay. In the musical training, they used the common music for this game: "Balearic Pumping, Natural, Rum n' Bass, and POP/STARS". In addition, they found that the harder level of the task required more mental effort, but musical training helped to improve cognitive load.

\subsubsection{Unnatural Music as an Auditory Feedback}
In Medvedev et al. study \cite{medvedev2015restorative}, aviation, motorcycle, and construction sounds are deliberated as unnatural sounds that negatively affect the participants’ perceptual and physiological strain. Participants generally considered these sounds as calming, restorative, and these sounds were compared with natural sounds, birds’ songs, and ocean waves. The findings of this study were that these sounds have an impact on greater physiological strain (higher skin conductance) and a slow rate of stress recovery. Moreover, they suggested that it may elevate the cognitive load. 

The researchers \cite{helps2014different} wanted to know whether the addition of white noise could affect children's thinking and attention with different levels of ability. They suggested, on the basis of the Moderate Brain Arousal model, every person has a different amount of natural neural noise, and improving the children’s brain activity may improve when adding external white noise. Mahmud et al. \cite{mahmud2024multimodal} also included white noise in their study, which was used in multimodal feedback methods for enhanced participant stability and their comfort.

\section{Methods}
\label{sec:methods}
This section outlines the experimental design, including system setup, participant details, auditory conditions, and procedures used to investigate the relationship between auditory feedback and cognitive load in VR navigation.

\subsection{System Description} 
\subsubsection{Balance Measurement}
In this study, we employed the BTrackS Balance Plate, which operates at a 25 Hz sampling frequency. This device was used to collect different types of balance, such as sway and stability, which helped to analyze the distribution of pressure applied to the plate. Ensuring high-resolution tracking of participants’ balance metrics data helped to identify and measure cognitive load for each scenario.
\subsubsection{Safety Equipment}
We utilized a harness system to ensure participants’ safety. When a participant conducts the VR task, there is a chance of falling, so to mitigate this risk, we used this system. 
\subsubsection{Computers, VR Equipment, and Software}
We had to set up the HTC VIVE Focus Vision, which is characterized by a 100-degree field of view, to ensure immersive and smooth visual experiences and collect the participant’s Eye tracking data. The Virtual Environment (VE) rendering and data collection were managed by using a high-performance computer that was configured with an Apple M3 chip, 8-core CPU, 10-core GPU, and RAM 8GB. Using the NI LabVIEW software (version 2020), which helps to collect the BTrackS Balance Plate data on our computer.
\subsubsection{Environment}
We performed our task in the eXtended Reality and Intelligence (XRei) Lab at Kennesaw State University’s College of Computing and Software Engineering, which was more than 1000 square feet in size. 
\subsubsection{Participants}
Firstly, we circulated information about participating in our pilot study. Then, we collected all demographic information from those who were interested. As we ran the pilot study, we recruited only six (06) participants with adults aged 18 and older from diverse backgrounds. The detailed information of the participants has been shown in the Table \ref{tab:participant info}.

\begin{table}[h!]
\centering
\caption{Descriptive statistics for participants}
\label{tab:participant info}
\resizebox{\columnwidth}{!}{%
\begin{tabular}{l cc cc cc cc}
\hline
\multicolumn{2}{c}{Participants} & \multicolumn{2}{c}{Age (years)} &
\multicolumn{2}{c}{Height (cm)} & \multicolumn{2}{c}{Weight (lb)} \\
 Male & Female & Mean & SD & Mean & SD & Mean & SD \\
\hline
 4 & 2 & 27 & 2.61 & 169.67 & 9.37 & 183 & 27.71 \\
\hline
\end{tabular}}
\end{table}

\subsection{Experimental Conditions}
\subsubsection{Auditory Feedback}
In this study, we utilized 15 directional auditory cues to provide auditory feedback that helps to investigate the cognitive load level while enhancing the balance and pupil diameter within VR-based navigation tasks. We were categorized into four types of audio (classical, natural, electronic, and unnatural) among 15 audio files. 

\subsubsection{No Feedback in VR}
To serve as a control condition, the participant played the task in VE without the involvement of any additional audio feedback. To maintain consistency in the environment, participants wore all the necessary equipment and adhered to other experimental conditions. This result was used as a baseline condition for comparison with all other conditions.
\begin{figure}[h]
    \centering
    \includegraphics[width=1\linewidth]{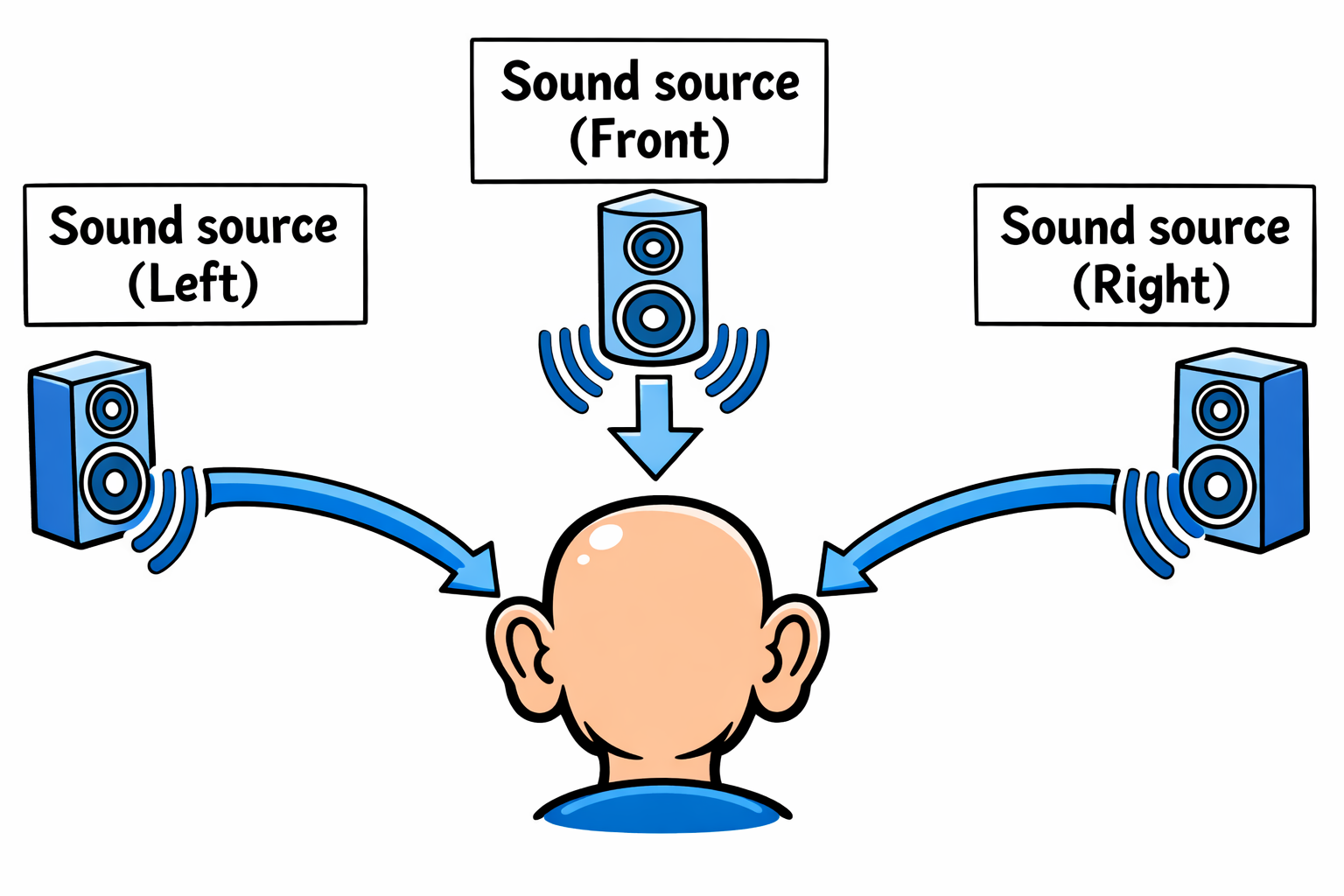}
    \caption{Auditory feedback system}
    \label{fig:auditory}
\end{figure}
\subsubsection{Auditory Feedback Setup}
To ensure precise direction-based three-dimensional sound localization, the sound was provided to the high-fidelity headphones. When a user explored the path and made the decision in which path he selected, our system provided a directional auditory cue. Auditory feedback in the left and right ears indicated to turn left and right, respectively, while the auditory feedback originating from the front indicated to move forward, like as Figure \ref{fig:auditory}. This feedback helped to guide which path was best suitable for the destination goal.

\subsection{Hypotheses}
Based on the literature analysis in Sections \ref{sec:introduction} and \ref{sec:background_study} and recent studies that discuss how extra auditory feedback systems can cause cognitive overload, we suggest the following hypothesis for this research:
\begin{itemize}
    \item H1: Our hypothesis is that without assistive feedback in VR, the cognitive load in navigation tasks can increase. When a participant performs a task without any feedback, it will increase the cognitive load \cite{paas1994instructional}.
    \item H2: We hypothesize that the participant's postural balance is related to the brain activity when he performs a task. Small et al. \cite{small2021influence} analyzed that when a participant performs a cognitive task while walking, their ability to maintain dynamic balance decreases under a higher cognitive load. 
    \item H3: Our hypothesis is that as the pupil diameter increases, it also indicates that the participant’s cognitive load is high. While a participant performs a task and makes a decision by taking more time, the pupil diameter is increased. Pupil diameter is more related to measuring or indicating cognitive load during a task \cite{kahneman1966pupil}. 
   \item H4: Participants exposed to classical and natural category audio will experience significantly lower cognitive load than when exposed to electronic and unnatural category audio.
    \item H5: There will be a significant difference in cognitive load across the different audio categories (classical, natural, electronic, and unnatural).

\end{itemize}

\subsection{Metrics}

\subsubsection{Mental Load Assessment}
A subjective rating system was used in our system to collect the mental load information provided by the participants. After each task completion, each participant was asked to fill out the NASA TLX form on a scale from 0 to 10. From this self-reported evaluation form, we could gauge the cognitive load with each condition. Using this numerical rating system allowed us to measure cognitive load and compare it with other performance metrics and conditions. 
\subsubsection{Center of Pressure (CoP) Velocity}
The term CoP velocity refers to the information about the dynamic shifts of the body’s center of pressure during the participant’s involvement in various tasks or conditions. For each condition, this metric was captured systematically and utilized to monitor and measure the changes in balance. After that, it helped to correlate with the participant's cognitive load.

\subsubsection{Simulator Sickness Questionnaire (SSQ)}
We utilized the SSQ in our system, which helps to analyze the prevalence of cybersickness and its potential impact on cognitive load and postural stability \cite{kennedy1993simulator}. To assess the physiological discomfort and symptoms, this SSQ was designed with 16 questions that were collected after the session. Our system analyzed this information to identify the participants’ cybersickness and also investigated how these symptoms are related to the cognitive load and balance performance. 

\subsection{Study Procedure} 

\subsubsection{Experimental Setup}
The study was approved by the Institutional Review Board (IRB). The participant wore the safety harness and headset to ensure stability and safety throughout the sessions, and placed the balance board in the center position of the safety harness. Participants’ balance was collected by using a balance board, and head movement and eye tracking data were stored using an HMD headset during all conditions. Balance signal, head movement, and eye tracking were all recorded as numerical measurements, which were processed for analysis.

\subsubsection{Pre-Session Questionnaires}
The Activities-specific Balance Confidence (ABC) questionnaire \cite{schepens2010short} and the SSQ \cite{kennedy1993simulator} were answered by the participants after finishing the whole session. Moreover, in each trial user rated the NASA TLX form on a scale from 0 to 10 for measuring the cognitive load.
\subsubsection{Experimental Tasks}
Participants performed a maze navigation task that was conducted in virtual environments. To control for potential order effects, the task order and navigation-oriented auditory feedback were randomized and counterbalanced.
\subsubsection{Baseline Measurements}
Participant conducted the maze navigation task without any auditory feedback and captured all information, such as eye tracking, head movement, balance, and the user's mental assessment data, as baseline measurements. This trial lasts about one minute. Finally, these baseline measurements helped to compare with all other conditions and decide whether to correlate with each other.
\subsubsection{Tasks in Virtual Reality}
The VR task is similar to the baseline activity, including the additional navigation-oriented auditory feedback conditions. These feedback conditions were randomized and counterbalanced, lasting around one minute.

\subsubsection{Maze Navigation in VR}
The VR Maze Game was chosen as the experimental setting because navigating a maze requires various cognitive functions, such as spatial reasoning, working memory, attention control, and decision-making, all of which are important to cognitive load. To navigate a maze, participants must constantly encode and remember spatial information, plan their paths, track their mistakes, and adjust to varying audio cues \cite{setu2024mazed}. All of these things naturally cause changes in mental workload that can be measured.

\begin{figure}[h]
    \centering
    \includegraphics[width=1\linewidth]{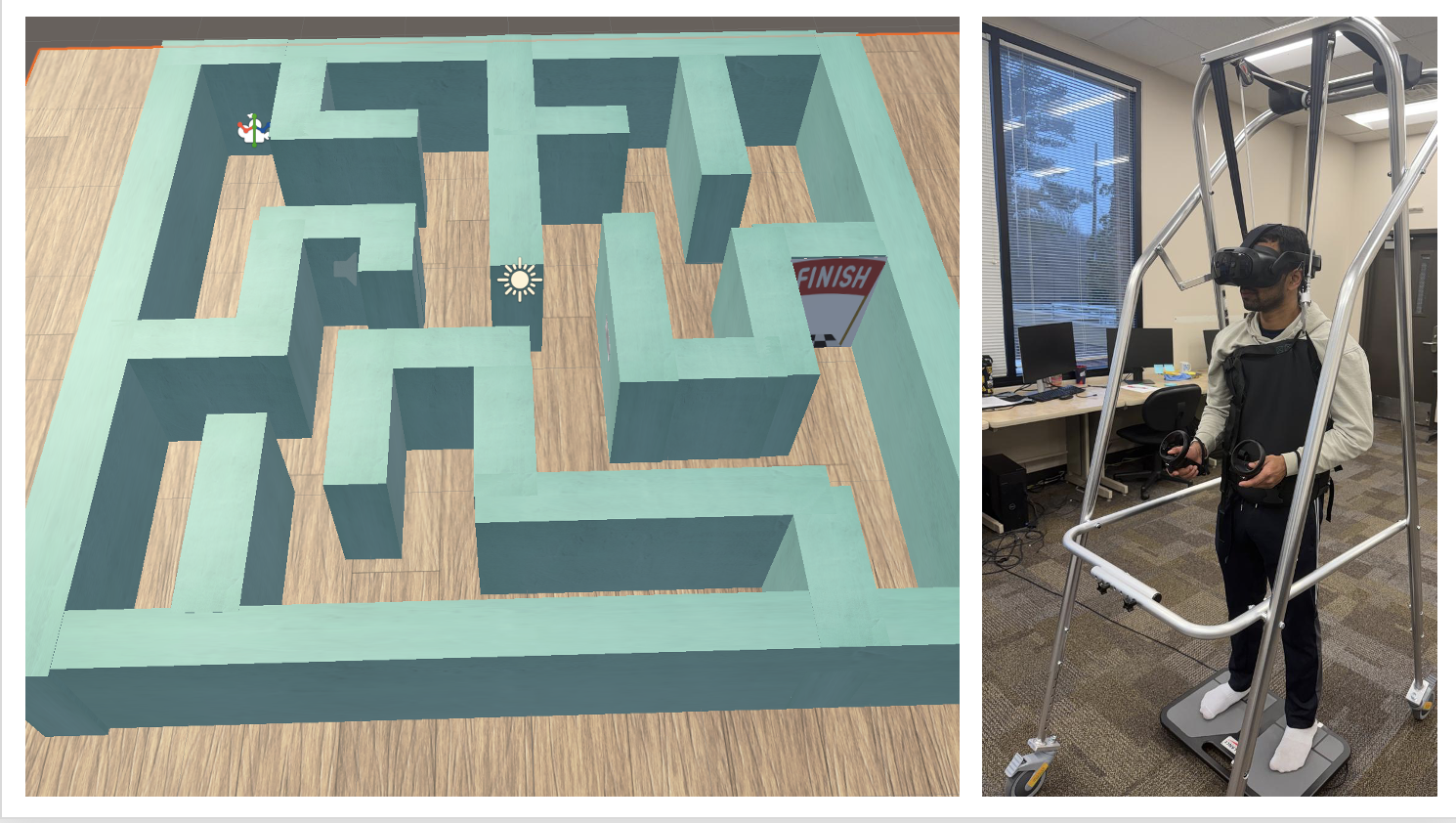}
    \caption{ a) Top-down view of the virtual maze environment. b) Participants navigating the virtual maze while standing on the balance board and wearing a harness for safety.}
    \label{fig:overview}
\end{figure}

The participants were placed in a flat space during the maze task, which is shown in the Figure \ref{fig:overview}. In this maze navigation task, the participant used the HTC VIVE Focus Vision headset. Using HTC Vive controllers, the participant moved and turned in the maze environment. The left controller was used for moving, and the right controller was used for turning in the path. When a user explored the path and made the decision in which path he selected, our system provided directional auditory cues. Auditory feedback in the left and right ears indicated to turn left and right, respectively, while the auditory feedback originating from the front indicated to move forward. This feedback helped to guide which path was best suitable for the destination goal. The entire session will be conducted based on navigation-oriented auditory feedback, and it took 1-2 minutes to navigate mazes.

\subsubsection{Post-Session Questionnaires}
After each condition, participants completed the NASA–TLX survey to measure cognitive load. Following the completion of all conditions, users were asked to fill out two distinct questionnaires: the SSQ and a demographic questionnaire. These questionnaires played a vital role in collecting all information related to simulator sickness experienced and the relevant background information on the users.

\section{Statistical Analysis \& Results}
\label{sec:stat_analysis}
This section presents the statistical methods applied to the collected data and discusses the key findings regarding cognitive load, balance, and pupil diameter across different auditory conditions.

For the statistical analysis, ANOVA and t-test were used in different studies \cite{mahmud2024multimodal, mahmud2023auditory}. We utilized a mixed model ANOVA for our experimental records. For this study, we designed a 3×5 factorial structure that evaluates the impact of three independent variables (balance, cognitive load, and pupil diameter), each with multiple participants, on the dependent variables. For further analysis of the ANOVA test, we performed pairwise comparisons using two-tailed t-tests. These comparisons were measured for differences between all combinations of study conditions. From these outcomes, we can determine statistical significance, thereby boosting the robustness of our findings.

\label{sec:results}
\begin{table*}[ht]
\centering
\caption{Correlation between balance, pupil diameter and cognitive load for various auditory conditions}
\label{tab:balvscog}
\renewcommand{\arraystretch}{1.4}
\resizebox{\textwidth}{!}{
\begin{tabular}{lccccccccccccc}
\hline
\textbf{Comparison} &
\textbf{$t_{CgLd}$} &
\textbf{$p_{CgLd}$} &
\textbf{$d_z^{CgLd}$} &
\textbf{$t_{Bal.}$} &
\textbf{$p_{Bal.}$} &
\textbf{$d_z^{Bal.}$} &
\textbf{$t_{Pupil}$} &
\textbf{$p_{Pupil}$} &
\textbf{$d_z^{Pupil}$} &
\textbf{$r_{(Bal,CgLd)}$} &
\textbf{$p_{r_{(Bal,CgLd)}}$} &
\textbf{$r_{(Pupil,CgLd)}$} &
\textbf{$p_{r_{(Pupil,CgLd)}}$}\\
\hline
Classical  \\vs No-Audio 
 & -2.47 & 0.0568 & -1.01 & -0.11 & 0.918 & -0.04  & -0.50 & 0.638 & -0.20
 & -0.82 & 0.0480$^*$ & 0.872 & 0.0236$^*$\\
\hline
Electronic \\vs No-Audio 
& -1.40 & 0.2209 & -0.57 & -0.17 & 0.874 & -0.07 & -1.04 & 0.347 & -0.42
& -0.47 & 0.3507 & 0.555 & 0.2532\\
\hline
Natural  \\vs No-Audio 
& -2.33 & 0.0676 & -0.95 & -0.46 & 0.666 & -0.19  & -0.62 & 0.564 & -0.25
& -0.98 & 0.00048$^*$ & 0.803 & 0.0543\\
\hline
Unnatural  \\ vs No-Audio 
& 0.16 & 0.875 & 0.07 & -0.06 & 0.954 & -0.02 & -2.75 & 0.0401$^*$ & -1.12
& -0.72 & 0.1072 & 0.766 & 0.0758\\
\hline
\\
\end{tabular}
}
\small
\textit{Note.} $t$ = paired-sample $t$ statistic; $p$ = probability value; $d_z$ = Cohen's effect size for paired samples; 
$r$ = Pearson correlation between balance, pupil diameter and cognitive load. $^*$ indicates $p < .05$.
\end{table*}
\subsection{Effect of Auditory Feedback on Cognitive Load}
The repeated-measures ANOVA provided a significant difference in the impact of different auditory feedback on cognitive load, F(4, 20) = 7.02, p = .0011. The result demonstrated that classical and natural audio feedback conditions reduced the cognitive load more than the no-audio feedback condition. Classical Music (Mean = 2.12) and Natural Music (Mean = 2.34) have lower cognitive value than Electronic Music (Mean = 2.89), and the highest is for Unnatural Music (Mean = 3.27). Thus, the result supported our hypothesis H5.

\subsection{Pairwise Comparisons Against No-Audio Baseline}
We showed in Table \ref{tab:balvscog} each sound condition with the no-audio baseline using a t-test. This analysis demonstrated that classical music (t=-2.47, p=.057, dz=-1.01) and natural music (t=-2.33, p=.068, dz=-0.95) presented marginally significant improvements in cognitive load. On the other side, Electronic Music (t=-1.40, p=.221, dz=-0.57) and Unnatural Music (t=0.16, p=.875, dz=0.07) had no meaningful difference compared to the baseline condition. This analysis established that classical and natural audio categories have a greater effect on cognitive load than other categories, thus supporting H4.  

\subsection{Relationship Between Cognitive Load and Balance}
\begin{figure}[h]
    \centering
    \includegraphics[width=1\linewidth]{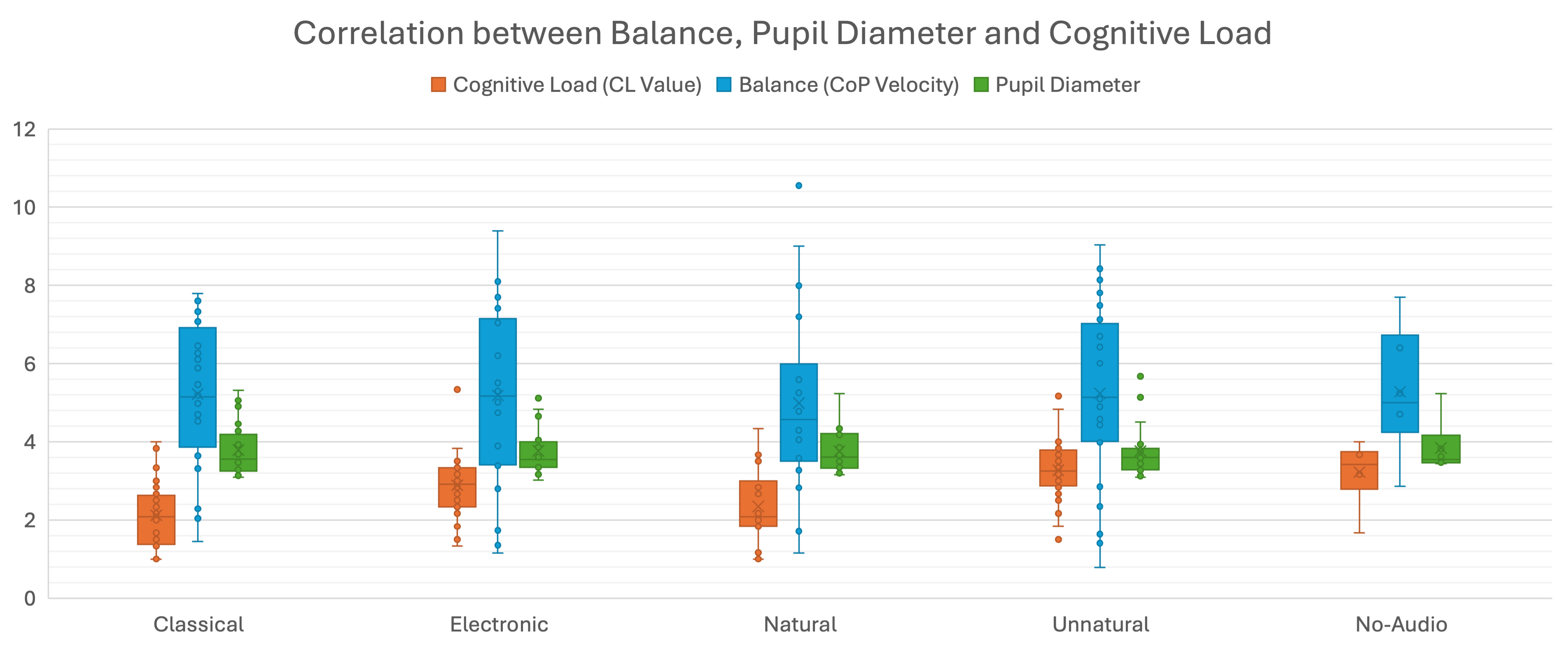}
    \caption{Correlation between Balance, Pupil Diameter and Cognitive Load for Different Types of Auditory Feedback}
    \label{fig:clvsblvspd}
\end{figure}

The correlation analysis in Table \ref{tab:balvscog} pointed out that cognitive load and postural balance have a reverse relationship. Classical Music (r=-0.82, p=.048) and Natural Music (r=-0.98, p$<$.001) have statistically significant results, while Electronic Music (r=-0.47, p=.351) and Unnatural Music (r=-0.72, p=.107) showed weaker and non-significant trends in Figure \ref{fig:clvsblvspd}. Therefore, it confirmed that a higher cognitive load is associated with lower postural balance, supporting H2.

\subsection{Relationship Between Cognitive Load and Pupil Diameter}
The table indicates a positive relation between pupil diameter and cognitive load across all auditory feedback conditions. For the classical category (r=0.872, p=0.0236), there is a statistically significant correlation, while natural music (r=0.803, p=0.0543) has a marginal relationship. On the other hand, electronic (r=0.555, p=0.2532) and unnatural categories (r=0.766, p=.0758) exhibited non-significant trends. Overall, these results indicated that higher cognitive load tends to be associated with increased pupil diameter in Figure \ref{fig:clvsblvspd}, which partially supported H3.  Though only the classical category fills the conventional levels of statistical significance.

\subsection{Assistive Auditory Feedback vs No-Audiotory Feedback }
From the comparison of cognitive load between no-audio feedback and different categories of assistive auditory feedback, the classical and natural categories provide better performance in Figure \ref{fig:clvsblvspd}. Since cognitive load was strongly related to balance, these reductions indirectly helped maintain postural stability. So, assistive auditory feedback can make it easier to navigate tasks in a VR environment that is supported by H1.

\subsection{Simulator Sickness Questionnaire}
We calculated a two-tailed t-test between pre-session SSQ score
and post-session SSQ score of participants. We found the statistically significant difference between pre-session SSQ and post-session SSQ of the participants. We got the result t(5) = 4.6377, p = .005644, Cohen's dz = 1.89 for participants.

\section{Discussion} 
\label{sec:discussion}
This section interprets the experimental results, highlighting their implications for cognitive load management, balance control, and the design of effective auditory feedback in VR systems.

\subsection{Effect of Assitive Auditory Feedback on Cognitive Load}
The result showed that the cognitive load in the VR navigation task is significantly influenced by using auditory feedback. Without auditory feedback, users only rely on visual scanning, which increases the mental workload. On the other hand, specifically classical and natural music categories represented lower cognitive loads than no-audio feedback and other audio categories. From this result, it supported our hypothesis (H1) that assistive auditory feedback can improve the cognitive load level due to providing external cues, which helps to make decisions easily during navigation tasks. 

\subsection{Cognitive Load–Balance Coupling in VR}
From the result analysis, we got the key finding that a strong negative correlation exists between cognitive load and balance. In the classical and natural auditory feedback conditions, the participants’ postural balance improved with lower cognitive load. This supports the dual-task interference theory, which suggests that the brain has limited attentional resources. When more attention is used for cognitive processing, less attention is available for motor control \cite{woollacott2002attention}. In VR environments, balance is already challenged by sensory conflict and spatial uncertainty. Therefore, reducing cognitive load through auditory feedback becomes especially important for maintaining balance. 

\subsection{Cognitive Load–Pupil Diameter Coupling in VR}
Another key finding of this study is that a strong positive relationship between cognitive load and pupil diameter was analyzed. For the classical and natural auditory feedback conditions, we found that higher cognitive load was associated with increased pupil diameter. These outcomes indicate that pupil diameter reflects less attention on cognitive resources.

\subsection{Design Implications for VR Navigation Systems}
This pilot study provides a clear overview of VR interface design for navigation tasks. It reveals that naturalistic, continuous, and low-complexity sounds are more comfortable and user-friendly than highly dynamic audio in assistive auditory navigation systems. Classical and natural music categories perform better in terms of cognitive load, pupil diameter, and balance. While designing VR environments, researchers and designers should consider such audio feedback as a fundamental accessibility feature, especially when considering older users or users with balance impairments.

\section{Conclusion and Future Work}
\label{sec:conclusion}
This section summarizes the main findings of the study and outlines potential directions for future research to further improve VR accessibility and user experience. This pilot study utilized an applicable approach for mitigating cognitive load, which is strongly associated with improved balance and pupil diameter, by using assistive directional auditory cues. We found that the classical and natural music categories were more effective than other categories. The addition of naturalistic audio feedback in VR environments helps users to become more efficient, safer, and more accessible, especially those who are vulnerable to cognitive overload or balance impairment. These results established that the selection of auditory cues is an important factor in designing VR systems. Moreover, the results also revealed the correlation between cognitive load, postural balance, and pupil diameter, which will motivate us to run a user study with many participants to investigate the interesting findings. The results will help validate the preliminary findings of this pilot study and provide stronger evidence for designing VR systems that incorporate effective auditory cues to improve user safety, efficiency, and accessibility.

\printbibliography

\end{document}